\documentclass[useAMS,usenatbib]{mn2e}
\usepackage{txfonts}
\newcommand{\be}{\begin{equation}}
\newcommand{\ee}{\end{equation}}
\newcommand{\bdm}{\begin{displaymath}}
\newcommand{\edm}{\end{displaymath}}
\def\dmf{\dot{\mathfrak{M}}}
\title[The origin of long-period X-ray pulsars]
{The origin of long-period X-ray pulsars}
\author[N.R.\,Ikhsanov]
{N.R. Ikhsanov$^{1,2}$\thanks{E-mail: ikhsanov@ast.cam.ac.uk}\\
$^1$Institute of Astronomy, University of Cambridge,
Madingley Road, Cambridge CB3\,0HA, UK\\
$^2$Central Astronomical Observatory of the Russian Academy of Sciences, Pulkovo
65--1, 196140 Saint-Petersburg, Russia}
\begin{document}
\date{Accepted ; Received }
\pagerange{\pageref{firstpage}--\pageref{lastpage}}
\pubyear{2006}
\maketitle
\begin{abstract}
Several relatively bright, persistent X-ray sources display regular pulses, with periods
in the range of $(0.7-10)\times 10^3$\,s. These sources are identified with massive close
binaries in which a neutron star accretes material onto its surface. The observed
pulsations in all of them, but one, are unambiguously associated with the spin period of
the neutron star. Analyzing possible history of these pulsars I conclude that the neutron
stars in these systems undergo spherical accretion and their evolutionary tracks in a
previous epoch contained three instead of two states, namely, {\it ejector}, {\it
supersonic propeller}, and {\it subsonic propeller}. An assumption about a supercritical
value of the initial magnetic field of the neutron stars within this scenario is not
necessary. Furthermore, I show that the scenario in which the neutron star in
2S\,0114+650 is assumed to be a magnetar descendant encounters major difficulties in
explaining the evolution of the massive companion. An alternative interpretation of the
spin evolution of the neutron star in this system is presented and the problem raised by
association of the $10^4$\,s pulsations with the spin period of the neutron star is
briefly discussed.

\end{abstract}

  \section{Introduction}

A newly formed neutron star is presumed to rotate rapidly with a period of a fraction of
a second. Its rotational rate then decreases, initially by the conventional spin-powered
pulsar energy-loss mechanism (ejector state), and later by means of the interaction
between the magnetosphere of the neutron star and the stellar wind of its companion
(propeller state). As the spin period of the neutron star reaches a critical value the
accretion of material onto its surface starts (accretor state) and the star switches on
as an X-ray pulsar \citep[for a review see, e.g.,][]{Bhattacharya-van-den-Heuvel-1991,
Iben-Tutukov-Yungelson-1995}.

This scenario has been numerically studied by Urpin, Konenkov \& Geppert (1998). Their
calculations have been performed under assumptions that the initial magnetic field of
the neutron star is subcritical (i.e. $B_0 < B_{\rm cr}=m_{\rm e}^2 c^3/e \hbar = 4.4
\times 10^{13}$\,G) and has a crustal origin. They have also assumed that the accretion
onto the stellar surface starts as soon as the centrifugal barrier at its magnetospheric
boundary becomes ineffective. The period at which this condition is satisfied is defined
by equating the corotational radius of the star, $r_{\rm c} = \left(GM P_{\rm s}^2/4
\pi^2\right)^{1/3}$, to its magnetospheric radius $r_{\rm m} = \kappa \left(\mu^2/\dmf
\sqrt{2GM}\right)^{2/7}$, that gives
      \be\label{pcd}
P_{\rm cd}(t_{\rm sd}) \simeq 18\ \kappa^{3/2}\ m^{-5/7}\ \left(\frac{\mu(t_{\rm
sd})}{10^{30}\,{\rm G\,cm^3}}\right)^{6/7} \left(\frac{\dmf(t_{\rm sd})}{10^{15}\,{\rm
g\,s^{-1}}}\right)^{-3/7}\ {\rm s},
   \ee
\citep[for a discussion see, e.g.,][]{Pringle-Rees-1972, Illarionov-Sunyaev-1975,
Stella-White-Rosner-1986}. Here $\kappa$ is the parameter accounting for the geometry of
the accretion flow, which ranges from 0.5 to 1 \citep{Ghosh-Lamb-1978}, and $t_{\rm sd}$
is the total spin-down time of the star in the ejector and propeller states. $M$,
$P_{\rm s}$, and $\mu$ are the mass, spin period, and the dipole magnetic moment of the
neutron star, respectively. $\dmf = \pi r_{\rm G}^2 \rho_{\infty} V_{\rm rel}$ is the
mass with which a neutron star interacts in a unit time as it moves through the wind of
a density $\rho_{\infty}$ with a velocity $V_{\rm rel} = \sqrt{V_{\rm ns}^2 + V_{\rm
w}^2}$, where $V_{\rm ns}$ is the linear velocity of the neutron star's orbital motion
and $V_{\rm w}$ is the wind velocity of its massive companion. The parameter $r_{\rm
G}=2GM/V_{\rm rel}^2$ denotes the Bondi radius of the neutron star, and
$m=M/1.4\,M_{\sun}$. The acceleration torque applied to the star in the accretor state
has been evaluated as $K_{\rm su} = \zeta \dmf (GM r_{\rm m})^{1/2}$, where $\zeta$ is
the efficiency parameter, which in the calculations has been normalized to 0.1. Finally,
the evolutionary tracks of neutron stars have been computed on a time scale $t_{\rm ms}
\sim 2 \times 10^6$\,yr, which corresponds to the life-time of a $30\,{\rm M_{\sun}}$
star on the main sequence \citep{Bhattacharya-van-den-Heuvel-1991}.

According to their results the spin period of a neutron star rapidly increases during
the ejector and propeller states to a maximum value $\sim P_{\rm cd} (t_{\rm sd})$ and
gradually decreases as the accretion of material onto their surface starts \citep[see
Figs.~2 and 4 in][]{Urpin-etal-1998}. The maximum period, which a neutron star with $B_0
< B_{\rm cr}$ is able to reach on a time scale $t_{\rm sd} < t_{\rm ms}$ under the above
conditions is limited to a few hundred seconds. In particular, for the case
$B_0=10^{13}$\,G, the longest period is $\sim 500$\,s, which the neutron star reaches at
$\dmf \sim 10^{-13}\,{\rm M_{\sun}\,yr^{-1}}$. The magnetic field of the neutron star at
the end of the propeller state is a factor of 3 weaker that its initial value, $B(t_{\rm
sd}) \sim 0.3 B_0$, and decreases by almost an order of magnitude on a time scale of
binary evolution, $B(t_{\rm ms}) \sim 0.1 B_0$.

The long-period pulsars within this scenario are expected to be relatively weak X-ray
sources. Indeed, solving Eq.~(\ref{pcd}) for $\dmf$ and putting the result to $L_{\rm X}
= \dmf GM/r_{\rm ns}$ one finds
 \be
L_{\rm X} \la 4 \times 10^{33}\ m^{-2/3}\ \left(\frac{P_{\rm s}}{500\,{\rm
s}}\right)^{-7/3} \left(\frac{B(t_{\rm sd})}{0.3 B_{\rm cr}}\right)^2 \left(\frac{r_{\rm
ns}}{10^6\,{\rm cm}}\right)^{-1}\ {\rm erg\,s^{-1}},
 \ee
where $r_{\rm ns}$ is the neutron star radius. The pulsars remain in the low-luminosity
state as long as the mass transfer is operated via the wind-fed mechanism. As the
massive companion overflows its Roche lobe the mass-transfer rate increases. However,
this transition will unavoidably be accompanied with a formation of an accretion disc
and, therefore, a rapid spin-up of the neutron star on a time scale of
\citep{Davidson-Ostriker-1973, Ghosh-Lamb-1978}
 \be\label{tsu}
\Delta t \simeq \frac{2 \pi I}{\dmf \sqrt{G M r_{\rm m}}} \left(\frac{1}{P_{\rm s}} -
\frac{1}{P_{\rm cd}(t_{\rm sd})}\right),
 \ee
where $I$ is the neutron star moment of inertia. Under the conditions of interest, i.e.
$I=10^{45}\,I_{45}\,{\rm g\,cm^2}$ and $\dmf = 10^{15}\,\dmf_{15}\,{\rm g\,s^{-1}}$, the
spin-up time scale is $\la 1000$\,yr. This indicates that the probability to detect the
pulsar in this stage is almost negligible.

There are, however, seven X-ray pulsars whose periods substantially exceed 500\,s (see
Tab.\,\ref{tab-1}). These sources are associated with massive close binaries in which a
magnetized neutron star accretes material onto its surface. All of them but one are
persistent X-ray emitters with an average luminosity $\ga 10^{34}\,{\rm erg\,s^{-1}}$,
which implies $\dmf \ga 10^{-12}\,{\rm M_{\sun}\,yr^{-1}}$. Observations of a cyclotron
line in the X-ray spectrum of X\,Per suggest that the surface field of the neutron star
is $3.3 \times 10^{12}$\,G \citep{Coburn-etal-2001}. There is some evidence \citep[see,
e.g.,][]{Li-van-den-Heuvel-1999} that the magnetic field of neutron stars in other
sources listed in Tab.\,\ref{tab-1} is of the same order of magnitude.

 \begin{table}
  \caption{Very long-period X-ray pulsars}
   \label{tab-1}
    \begin{tabular}{@{}lccccc}
    \hline
    Name & Sp. type & $P_{\rm s}$, s & $P_{\rm orb}$, d & $\log{L_{\rm x}}^*$ &  Ref.$^{**}$    \\
     \hline
    J170006-4157 & --     & 715 & --  & 34.7 & [1] \\
    0352+309     & B0~Ve  & 837  & 250 & 34.7 -- 35.5 & [2, 3]\\
    (X Per)      & (O9.5~IIIe)   &      &     &  &  \\
    J1037.5-5647$^{\dag}$ & B0~V-IIIe & 862  &  -- & 34-35  & [4, 5]\\
    J2239.3+6116 & B0~Ve & 1247 & 262.2 & 33 -- 36 & [6, 7, 8] \\
                 & (B2~IIIe) & & & & \\
    J0103.6-7201 & O5~Ve & 1323 & -- & 35.3 -- 36.8 & [9] \\
    J0146.9+6121 & B1~III-Ve & 1412 & -- & 34.6 -- 36 & [10] \\
    (V831\,Cas)  &           &      &           &  \\
    2S\,0114+650 & B1~Ia & 10008$^{\ddag}$ & 11.6 & 35.7 -- 36 & [11, 12]\\
    \hline
   \end{tabular}

   \medskip
   $^*$\,$L_{\rm x}$ is expressed in erg\,s$^{-1}$,\\
   $^{**}$\,References to the table:
   [1]~\citet{Torii-etal-1999};
   [2]~\citet{Delgado-Marti-etal-2001};
   [3]~\citet{Haberl-etal-1998};
   [4]~\citet{Reig-Roche-1999};
   [5]~\citet{Motch-etal-1997};
   [6]~\citet{in-'t-Zand-etal-2000};
   [7]~\citet{in-'t-Zand-etal-2001};
   [8]~\citet{Ziolkowski-2002};
   [9]~\citet{Haberl-Pitsch-2005};
   [10]~\citet{Haberl-Sasaki-2000};
   [11]~\citet{Reig-etal-1996};
   [12]~\citet{Hall-etal-2000}.\\
   $^{\dag}$~The Be/X-ray transient pulsar.\\
   $^{\ddag}$~An association of the 10008\,s pulsations with the spin period of the neutron star
   has recently been challenged by \citet{Koenigsberger-etal-2006}, see text.
  \end{table}

The origin of these sources within the above scenario is rather unclear. Indeed, the
value of $P_{\rm cd}(t_{\rm sd})$ for $B(t_{\rm sd}) \simeq 0.3 B_{\rm cr}$ and $\dmf
\ga 10^{-12}\,{\rm M_{\sun}\,yr^{-1}}$ is limited to $< 300$\,s. For a neutron star with
$B_0 \leq B_{\rm cr}$ to reach a period of $\sim 10^3$\,s the strength of the wind
should be $\leq 6 \times 10^{-14}\,{\rm M_{\sun}\,yr^{-1}}$. In this case, however, the
luminosity of the pulsar during the wind-fed accretor state is $\la 7 \times
10^{32}\,{\rm erg\,s^{-1}}$, i.e. by a factor of 130 smaller than the average X-ray
luminosity of the sources listed in Tab.\,\ref{tab-1}. On the other hand, a possibility
that the massive star overflows its Roche lobe at least in the case of X~Per,
J2239.3+6116, and 2S\,0114+650 can be excluded. Hence, the only way to explain the
origin of the long-period pulsars within the above scenario is to invoke a rather
controversial assumption that the mass-loss rate of their massive companions has
recently been increased by more than two orders of magnitude. Finally, the $10^4$\,s
spin period of neutron star in 2S\,0114+650 represents an exceptional case which within
the above scenario appears to be a miracle.

A possibility to improve the situation by assuming that the neutron star in 2S\,0114+650
was born as a magnetar \citep[i.e. $B_0 \gg B_{\rm cr}$, see,
e.g.,][]{Thompson-Duncan-1993} has been discussed by \citet{Li-van-den-Heuvel-1999}.
They have shown that the spin-down time scale of a star with $B_0 \sim 10^{15}$\,G to a
period $\sim 10^4$\,s under the condition $\dmf \la 10^{14}\,{\rm g\,s^{-1}}$ is close
to $10^5$\,yr, which is comparable with the characteristic time of magnetic field decay
of magnetars \citep{Colpi-etal-2000}. Following this finding they have suggested a
scenario in which the neutron star in 2S\,0114+650 is assumed to be a magnetar
descendant undergoing a spherical accretion with a very low angular momentum transfer
rate. They have also pointed out that this scenario, being applied to other long-period
pulsars, may shed a new light to the origin of strong magnetic field of neutron stars in
such systems as A0535+26, Vela\,X-1, GX\,1+4, 4U\,1907+09, 4U\,1538-52, and GX\,301-2.

It appears, however, that the rotational rate of a neutron star, which forms during the
first supernova explosion in a massive binary system, is insufficient for its magnetic
field to be amplified over the critical value \citep{Thompson-Murray-2001,
Heger-etal-2003, Petrovic-etal-2005}. It is more likely, that magnetars form at the
latest stages of binary evolution, namely, during the second supernova explosion or, the
most probably, during a coalescence of two neutron stars \citep[see, e.g.,][and
references therein]{Price-Rosswog-2006}. In this case, however, the fraction of
magnetars in binary systems can unlikely exceed 1 per cent and most of them are expected
to have a black hole companion \citep[for a discussion see,
e.g.,][]{Popov-Prokhorov-2006}. Moreover, the kicks which magnetars are expected to get
during their formation are too large for the binary system to survive \citep[][and
references therein]{Wheeler-etal-2000}. In this light, a probability for a magnetar to
be accompanied with a massive main sequence star appears to be almost negligible.

While these are not compelling arguments against the long-period pulsars to be
descendants of magnetars, they suggest that alternative possibilities to solve the
problem might be more fruitful. One of them is discussed in this paper. Namely, I show
that the spin periods of long-period pulsars can be explained in terms of evolutionary
tracks constructed by \citet{Davies-etal-1979} and \citet{Davies-Pringle-1981} and
recently improved by \citet{Ikhsanov-2001a} provided the neutron stars in these systems
undergo spherical accretion. These tracks contain an additional, the so called subsonic
propeller, state, which has not been taken into account in the models considered by
\citet{Urpin-etal-1998} and \citet{Li-van-den-Heuvel-1999}. The spin period of a neutron
star during this state increases in a relatively short time, $\tau_{\rm d} \ll t_{\rm
sd}$, to a value expressed by Eq.~(\ref{pbr}), which under the conditions of interest
lies in the interval $10^3-10^4$\,s. The assumption about the supercritical value of the
initial magnetic field of neutron stars for the interpretation of long-period pulsars
within this scenario is not required.

   \section{Accretion flow geometry}\label{flow-geometry}

The state of neutron stars in the long-period pulsars is unambiguously identified with
an accretor. The equation governing the spin evolution of stars in this state reads
\citep{Davidson-Ostriker-1973, Ghosh-Lamb-1978}
  \be\label{main}
2 \pi I \frac{d}{dt} \frac{1}{P_{\rm s}} = K_{\rm su} + K_{\rm sd},
 \ee
where $K_{\rm su}$ and $K_{\rm sd}$ are the acceleration and deceleration torques
applied to the neutron star, respectively. The deceleration torque is associated with
interaction between the stellar magnetic field and material situated at a distance $r
\geq r_{\rm c}$. The average value of this torque does not significantly depend on the
geometry of the accretion flow beyond the magnetosphere and in the general case can be
expressed as \citep{Lynden-Bell-Pringle-1974, Wang-1981, Lipunov-1992}
 \be\label{ksd}
K_{\rm sd} = - k_{\rm t}\ \mu^2/r_{\rm c}^3,
 \ee
where $(k_{\rm t}\la 1)$ is the dimensionless parameter of the order of unity. In
contrast, the acceleration torque is very sensitive to the flow geometry and varies from
its maximum value, $K_{\rm su}^{\rm d} \simeq \dmf \sqrt{GMr_{\rm m}}$, in the case of a
disc, to a significantly smaller value,
 \be\label{ksd}
K_{\rm su}^{\rm sph} \simeq \ \frac{1}{4}\ \xi\ \dmf\ \Omega_{\rm orb} r_{\rm G}^2,
 \ee
in the case of a spherical accretion \citep{Davidson-Ostriker-1973, Wang-1981}. Here
$\Omega_{\rm orb}=2 \pi/P_{\rm orb}$ is the orbital angular velocity and $P_{\rm orb}$
is the system orbital period. The parameter $\xi$ is the factor by which the angular
momentum accretion rate is reduced due to inhomogeneities (the velocity and density
gradients) in the accretion flow. Numerical simulations \citep[see,
e.g.,][]{Anzer-etal-1987, Taam-Fryxell-1988, Ruffert-1999} suggest that the average
value of this parameter is $<\xi>=0.2$.

As follows from Eq.~(\ref{main}), the spin period of an accreting neutron star evolves
to the so called equilibrium period, which is defined by setting $K_{\rm su}=K_{\rm
sd}$. The values of the equilibrium period in the case of a disc and spherical
accretion, respectively, are
 \be\label{peqd}
P_{\rm eq}^{\rm d} \simeq 18\ {\rm s}\ \times\ \kappa^{-1/4}\ k_{\rm t}^{1/2}\ m^{-5/7}\
\dmf_{15}^{-3/7}\ \left(\frac{B}{3.3 \times 10^{12}\,{\rm G}}\right)^{6/7},
 \ee
and
  \be\label{peqsph}
P_{\rm eq}^{\rm sph} \simeq 910\ {\rm s}\ \times\ k_{\rm t}^{1/2}\ \xi_{0.2}^{-1/2}\
m^{-3/2}\ \dmf_{15}^{-1/2}\ \left(\frac{B}{3.3 \times 10^{12}\,{\rm G}}\right)\ \times
 \ee
 \bdm
\times\ \left(\frac{V_{\rm rel}}{400\,{\rm km\,s^{-1}}}\right)^2\ \left(\frac{P_{\rm
orb}}{250\,{\rm d}}\right)^{1/2},
 \edm
where $P_{250}=P_{\rm orb}/250$\,days, and $\xi_{0.2}=\xi/0.2$. The normalization of
parameters in these equations is appropriate for X~Per, which is the best studied
persistent long-period pulsar.

As easy to see, the value of $P_{\rm eq}^{\rm d}$ under the conditions of interest is
much smaller than the spin period of neutron stars in the pulsars listed in
Tab.\,\ref{tab-1}. Hence, if a persistent accretion disc in these systems existed the
acceleration torque applied to the neutron star would significantly exceed the
deceleration torque. The neutron star in this case would regularly spinning-up at a rate
$\dot{P} \simeq P_{\rm s}^2 \dmf \sqrt{GMr_{\rm m}}/2 \pi I \sim - 5 \times
10^{-8}\,{\rm s\,s^{-1}}$, which implies the lifetime of the long-period pulsars to be
$< 1000$\,yr (see Eq.~\ref{tsu}). In particular, if the neutron star in 2S\,0114+650
accreted material from a disc its spin period would substantially increase on a time
scale of only 100\,yr, which is 4--5 orders of magnitude smaller than the average
lifetime of accretors in massive stars \citep{Urpin-etal-1998}. Thus, a probability to
observe neutron stars accreting from a disc at a rate $\sim 10^{15}\,{\rm g\,s^{-1}}$
during their long-period stage is almost negligible.

Furthermore, observations of X-ray source in X~Per show no evidence for a regular
spin-up of this pulsar. Instead, it exhibits apparent erratic pulse frequency variations
on a time scale of a few days, which are superposed with a 10--20\,yr spin-up/spin-down
trends around the average period of 837\,s \citep[][and references therein]{Haberl-1994,
Delgado-Marti-etal-2001}. The transitions between the spin-up and spin-down stages occur
without any significant variations of the system X-ray luminosity, and therefore, cannot
be associated with the transitions of the neutron star between the accretor and
propeller states. The spin-down events in this case could occur only if the value of
acceleration torque applied to this star is much smaller than $K_{\rm su}^{\rm d}$,
which argues against the presence of a persistent accretion disc in this system.

In contrast, the assumption about spherical geometry of the accretion flow allows us to
interpret the spin period of the neutron star in X~Per in terms of the equilibrium
period, $P_{\rm eq}^{\rm sph}$, provided the average (on a time scale of $> 20$\,yr)
value of the wind velocity is $350-400\,{\rm km\,s^{-1}}$. The spin-up/spin-down
behaviour of the source within this scenario can be associated with apparent variations
of the stellar wind velocity (e.g., due to activity of the massive component and the
orbital motion of the neutron star whose trajectory is inclined to the plane of
decretion disc of the Be-companion) and, possibly, the flip-flop instability of the
accretion flow \citep[see, e.g.,][]{Taam-Fryxell-1988}. The spherical accretion model is
also effective for the interpretation of other long-period pulsars provided their
orbital periods are of the same order of magnitude as that of X~Per, and the average
wind velocity is $\sim 400-800\,{\rm km\,s^{-1}}$.

The only exception is 2S\,0114+650. The condition $P_{\rm s} = P_{\rm eq}^{\rm sph}$ in
the case of this source implies $V_{\rm rel} \ga 3000\,{\rm km\,s^{-1}}$, which by a
factor of 2--3 exceeds the upper limits to the stellar wind velocity of B-type stars
inferred from the UV observations \citep[$800-1500\,{\rm km\,s^{-1}}$, see,
e.g.,][]{Bernacca-Bianchi-1981, Snow-1981, Marlborough-1982}. On the other hand, assuming
that the wind velocity of the massive companion is limited to $\la 1500\,{\rm
km\,s^{-1}}$ one finds $P_{\rm eq}^{\rm sph} \la 26$\,minutes. This indicates that either
the neutron star is a very young accretor and its period significantly exceeds the
equilibrium one, or the value of the parameter $\xi$ in the particular case of this
system is significantly smaller than its average value inferred from the numerical
simulations (see above). However, it cannot be excluded that this apparent discrepancy
occur because of a mistaken association of the $10^4$\,s pulsations with the spin period
of the neutron star. As recently shown by \citet{Koenigsberger-etal-2006}, these
pulsations may reflect a modulation of the B-supergiant wind caused by tidal interaction
between the non-synchronously rotating binary components. As the material of the wind is
captured by the neutron star and is accreted onto its surface the X-ray luminosity of the
source would be modulated with the same period, which under the conditions of interest
lies within the interval of 2--3\,hours. This scenario weakens the association of
$10^4$\,s pulsations with the spin period of the neutron star and, therefore, leaves the
question about the rotational rate of the star open \citep[for a discussion
see,][]{Koenigsberger-etal-2006}. As mentioned above, the condition $P_{\rm s} = P_{\rm
eq}^{\rm sph}$ for reasonable values of $V_{\rm rel}$ predicts $P_{\rm s} \sim
10-26$\,minutes. In this light, it is interesting to note that a detection of $\sim
15$\,minutes pulsations has been reported by \citet{Koenigsberger-etal-1983} and
\citet{Yamauchi-etal-1990}. These pulsations, however, have not been found in later X-ray
observations of this system.

Summarizing this section I can conclude that a presence of accretion disc in the
long-period pulsars is very unlikely. The neutron stars in these systems are, therefore,
undergoing spherical accretion.

  \section{Subsonic propeller}\label{subprop}

Let us now address the main question, namely, what is the maximum period to which a
neutron star undergoing spherical accretion can be spun-down on a time scale $t < t_{\rm
ms}$\,? A comprehensive analysis of this question has been first presented by
\citet{Davies-etal-1979} and \citet{Davies-Pringle-1981}. As they have shown the
spherically accreting neutron star in the state of propeller is surrounded by a hot
quasi-stationary turbulent atmosphere. The atmosphere forms as soon as the pressure of
relativistic wind ejected by the neutron star in the spin-powered pulsar state can no
longer balance the pressure of the surrounding material, and the latter, penetrating to
within the accretion radius of the star, interacts with the stellar magnetic field. This
interaction leads to formation of the magnetosphere with the equatorial radius $r_{\rm
m}$, which is defined by equating the ram pressure of the flow with the magnetic
pressure due to dipole field of the neutron star. The kinetic energy of the flow at the
magnetospheric boundary is converted into its thermal energy in the adiabatic shock. The
temperature of material in the shock increases to the adiabatic (or the so called
free-fall) temperature, $T_{\rm ff} = GMm_{\rm p}/k_{\rm B}r$, where $m_{\rm p}$ and
$k_{\rm B}$ are the proton mass and Boltzmann constant, respectively. The heated gas
expands with the velocity $V_{\rm ff} = (2GM/r)^{1/2}$, which significantly exceeds the
sound speed in the material captured by the star. The gas expansion is, therefore,
leading to a formation of the back-flowing shock, which propagates through the flow and
heats it up to the adiabatic temperature. Under the condition $\dmf \la \dmf_{\rm cr}
\simeq 2 \times 10^{18}\ {\rm g\,s^{-1}}\ m\ (V_{\rm rel}/10^8\,{\rm cm\,s^{-1}})$
\citep{Ikhsanov-2002} the atmosphere is extended up to the Bondi radius of the star,
$r_{\rm G}$, at which the thermal pressure of the heated material is equal to the ram
pressure of the surrounding gas, $\sim \rho_{\infty} V_{\rm rel}^2$. The formation of
the atmosphere prevents the stellar wind from penetrating to within the Bondi radius of
the neutron star. As the neutron star moves through the wind of its companion the
surrounding gas overflow the outer edge of the atmosphere with the rate $\dmf$ and the
mass of the atmosphere is conserved.

As long as the magnetospheric radius of the star exceeds its corotational radius (which
is equivalent to the condition $P_{\rm s} \la P_{\rm cd}(t_{\rm sd})$, see Introduction)
the linear velocity of the magnetosphere (which is assumed to co-rotate with the star)
at the radius $r_{\rm m}$ exceeds the sound speed in the surrounding material \citep[the
corresponding state is usually referred to as {\it supersonic
propeller,}][]{Davies-Pringle-1981}. The heating rate of the atmosphere due to the
propeller action by the star in this case significantly exceeds the rate of plasma
cooling due to the bremsstrahlung emission and turbulent motions \citep[for a discussion
see also,][and references therein]{Lamb-etal-1977}. The atmosphere during this state
remains hot, $T(r) \simeq T_{\rm ff}(r)$. The rotational energy loss by the neutron star
is convected up through the atmosphere by the turbulent motions and lost through its
outer boundary. The spin-down rate of the star in this state \citep[see Table in
][]{Davies-etal-1979} is the same as that evaluated by \citet{Illarionov-Sunyaev-1975}
and used by \citet{Urpin-etal-1998} in their calculations. This indicates that the
evolution of the spin and the magnetic field of a spherically accreting neutron star up
to a moment when its period reaches $P_{\rm cd}(t_{\rm sd})$ can be treated within the
results presented by \citet{Urpin-etal-1998}. In particular, this suggests that the
surface field of the star to the end of the supersonic propeller state is only by a
factor of 3 weaker than the initial field.

As the spin period exceeds $P_{\rm cd}(t_{\rm sd})$, the linear velocity of the
magnetosphere at the radius $r_{\rm m}$ becomes smaller than the sound speed
corresponding to the temperature $T_{\rm ff}(r_{\rm m})$. However, the effective
acceleration applied to the hot gas moving along the curved field lines,
    \be\label{eff}
g_{\rm eff}\ = \frac{G M}{r_{\rm m}^{2}(\theta)} \cos{\theta} - \frac{V_{\rm T_{\rm
i}}^{2}(r_{\rm m})}{r_{\rm curv}(\theta)},
   \ee
is directed outwards from the star as long as the temperature of the material situated
over the boundary is \citep{Elsner-Lamb-1976, Arons-Lea-1976}
 \be\label{tcr}
T_{\rm p}(r_{\rm m}) > T_{\rm cr} \simeq 0.1 T_{\rm ff}(r_{\rm m}).
 \ee
Here $r_{\rm curv}$ is the curvature radius of the field lines, $\theta$ is the angle
between the radius vector and the outward normal to the magnetospheric boundary and
$V_{\rm T_{\rm i}}(r_{\rm m})$ is the ion thermal velocity in the material located at
the inner edge of the atmosphere. Furthermore, the magnetospheric boundary under the
condition~(\ref{tcr}) is stable with respect to interchange instabilities and the plasma
penetration rate into the magnetosphere is limited to the rate of Bohm diffusion
\citep{Elsner-Lamb-1984}, which under the conditions of interest is
\citep{Ikhsanov-Pustilnik-1996}
   \be\label{dotmdif}
\dmf_{\rm B} \la 10^{11}\,{\rm g\,s^{-1}}\ \alpha_{0.1}^{1/2}\ m^{1/7}\
\left(\frac{B(t_{\rm sd})}{0.3 B_{\rm cr}}\right)^{-1/14}
 \left(\frac{\dmf}{10^{15}\,{\rm g\,s^{-1}}}\right)^{11/14}.
  \ee
Here $\alpha_{0.1}=\alpha/0.1$ is the diffusion efficiency, which is normalized
following \citet{Gosling-etal-1991}. Thus, the condition $r_{\rm m} < r_{\rm c}$ appears
to be necessary, but not sufficient for a direct accretion with the rate of $\sim \dmf$
onto the stellar surface to start. In addition, it is required that the cooling of
plasma at the base of the atmosphere is more effective than the heating.

As shown by \citet{Davies-Pringle-1981}, the cooling of the atmospheric plasma is
governed by the bremsstrahlung radiation and the convective motion. For these processes
to dominate the energy input into the atmosphere due to the propeller action by the
star, the spin period of the star should be $P_{\rm s} \ga P_{\rm br}$, where $P_{\rm
br}$ is the so-called break period, which is \citep{Ikhsanov-2001a}
   \be\label{pbr}
P_{\rm br}(t_{\rm sd}) \simeq\ 2000\ m^{-4/21}\ \left(\frac{B(t_{\rm sd})}{0.3 B_{\rm
cr}}\right)^{16/21}\ \left(\frac{\dmf}{10^{15}\,{\rm g\,s^{-1}}}\right)^{-5/7}\ {\rm s}.
      \ee
Under the conditions of interest, the break period significantly exceeds $P_{\rm
cd}(t_{\rm sd})$. This means that a spherically accreting neutron star is able to switch
its state from the {\it supersonic propeller} to {\it accretor} only via an additional
intermediate state, which is called the {\it subsonic propeller}. The star during this
state remains surrounded by the hot adiabatic atmosphere and its spin period increases
from $P_{\rm cd}$ to $P_{\rm br}$ on a time scale of \citep{Davies-Pringle-1981}
 \be\label{taud}
\tau_{\rm d} \simeq 4 \times 10^3\ I_{45}\ m^{2/7}\ \left(\frac{B(t_{\rm sd})}{0.3
B_{\rm cr}}\right)^{-8/7}\ \left(\frac{\dmf}{10^{15}\,{\rm g\,s^{-1}}}\right)^{-3/7}\
{\rm yr}.
 \ee
The accretion luminosity of the star during this state is limited to $\la \dmf_{\rm
B}GM/r_{\rm ns}$, which under the conditions of interest is $\la 10^{31}\,{\rm
erg\,s^{-1}}$. Finally, the surface magnetic field of the star in the subsonic propeller
state would not evolve significantly since $\tau_{\rm d} \ll t_{\rm sd}$.

   \section{Long-period pulsars}\label{lpp}

The results of incorporation of the subsonic propeller state into the evolutionary tracks
presented by \citet{Urpin-etal-1998} are shown in Tab\,\ref{tab-2} (the corresponding
evolutionary scenario will be further referred to as DFP-scenario). The second column of
the Table shows the ratio of the average (on a time scale of $t \ga \tau_{\rm d}$)
strength of the stellar wind during the spin-down epoch, $<\dmf>$, to the average mass
capture rate by the neutron star inferred from the X-ray luminosity, $<\dmf_{\rm x}> =
r_{\rm ns} <L_{\rm x}>/GM$. The third column represents the ratio of the initial surface
field of the neutron star to the critical field.

The long-period pulsars in Table\,\ref{tab-2} are divided into three groups. The first
group contains 4 sources whose origin within DFP-scenario can be explained provided the
average strength of the wind during the system evolution has not been changed
significantly and that the initial magnetic field of the neutron stars was subcritical.
The value of the ratio $B_0/B_{\rm cr}$ for these sources has been evaluated from
Eq.~(\ref{pbr}) by setting $<\dmf>/<\dmf_{\rm x}> =1$ and taking into account that
$B(t_{\rm sd}) \simeq 0.3 B_0$.

An application of DFP-scenario to J0103.6-7201 and J0146.9+6121 (the second group) leads
us to a conclusion that either the wind during a previous epoch was by a factor of 2--5
weaker than that inferred from the X-ray luminosity of these sources or the value of
$B(t_{\rm sd})$ was a factor of 2 larger than the average one estimated by
\citet{Urpin-etal-1998}, or both. According to the results reported by
\citet{Vink-de-Koter-Lamers-2000} and \citet{Fullerton-Massa-Prinja-2006} it is not
unusual for massive stars to lose material with a higher rate at later stages of their
evolution. It also cannot be excluded that the X-ray luminosity of these sources is
slightly overestimated or has been measured during a high state of activity of the
massive companions, which usually lasts 3--20\,years. In this light, the assumption
about a weaker wind in a previous epoch of J0103.6-7201 and J0146.9+6121 seems to be
rather reasonable. On the other hand, the duration of the spin-down epoch of neutron
stars with the initial field $\sim B_{\rm cr}$ in the wind of $\sim 10^{16}\,{\rm
g\,s^{-1}}$ is by a factor of 2-3 shorter than the average one evaluated by
\citet{Urpin-etal-1998}. This indicates that the assumption about a higher value of
$B(t_{\rm sd})$ has curtain theoretical grounds as well. However, both of these
arguments are not compelling to exclude a possibility that the neutron stars in
J0103.6-7201 and J0146.9+6121 were born as magnetars.

The third `group' is represented by only one (but the most puzzling) source
2S\,0114+650. The origin of this source within DFP-scenario can be explained provided
the strength of the wind in a previous epoch was by a factor of $\sim 30$ smaller than
the mass capture rate by the neutron star inferred from the observed X-ray luminosity.
Indeed, for the condition $P_{\rm br} = 10^4$\,s in the case $B_0 \leq B_{\rm cr}$ to be
satisfied the strength of the wind should be $\dmf \leq 10^{14}\,{\rm g\,s^{-1}}$. The
neutron star under these conditions is able to reach a period of $10^4$\,s within a time
$t_{\rm sd} < t_{\rm ms}$ (see Fig.\,2 in Urpin et~al. 1998, and Eq.~\ref{taud}).

On the other hand, an attempt to interpret this source within the scenario of
\citet{Li-van-den-Heuvel-1999} leads us to a significantly stronger limitation to the
value of $\dmf$. Indeed, the neutron star within their scenario is presumed to switch to
the accretor state as soon as its spin period reaches $P_{\rm cd}(t_{\rm sd})$. Setting
$B_0 \sim 10^{15}$\,G they have evaluated the spin-down time scale as $t_{\rm sd} \la
10^5$\,yr and, correspondingly, $B(t_{\rm sd}) \sim (1-4) \times 10^{14}$\,G, which
implies $\mu(t_{\rm sd}) \sim 10^{32}~\mu_{32}\,{\rm G\,cm^3}$. Putting this value to
Eq.~(\ref{pcd}) and solving the equation for $\dmf(t_{\rm sd})$ one finds $\dmf(t_{\rm
sd}) \la 2 \times 10^{13}~{\rm g\,s^{-1}} \times \kappa^{7/2} m^{-5/3} \mu_{32}^2
\left(P_{\rm cd}(t_{\rm sd})/10^4\,{\rm s}\right)^{-7/3}$. This means that the
interpretation of the long spin period of the neutron star in 2S\,0114+650 within the
magnetar hypothesis requires the strength of the wind in a previous epoch to be by a
factor of 250 smaller than that inferred from the X-ray luminosity of the source.

One of the ways to improve the situation is to assume that the evolutionary tracks of
magnetars contain the subsonic propeller state. In this case the neutron star can
comfortably reach a period of $10^4$\,s even if the strength of the wind is $\sim
10^{15}\,{\rm g\,s^{-1}}$ (see Sect.\,\ref{discussion}). However, a necessity to invoke
this assumption is not obvious. According to the evolutionary tracks reported by
\citet{Meynet-etal-1994}, a possible progenitor of the B1-supergiant was a O9.5~V star.
The typical mass-loss rate of this type stars \citep[see, e.g.,][and references
therein]{Crowther-Lennon-Walborn-2006, Fullerton-Massa-Prinja-2006} is by a factor of
10--50 smaller than the mass-loss rate of the massive companion in 2S\,0114+650 inferred
from the optical observations \citep{Aab-Bychkova-1984, Reig-etal-1996}. This indicates
that the value of the ratio $<\dmf>/<\dmf_{\rm x}>$, given in Tab.\,\ref{tab-2} has good
observational grounds and, therefore, allows us to interpret this object in terms of
DFP-scenario assuming the initial magnetic field of the neutron star to be subcritical.
Finally, I would like to remind that the association of the $10^4$\,s pulsations with the
spin period of the neutron star cannot be considered as finally justified (see
Sect.\,\ref{flow-geometry}). If the spin period of the neutron star is indeed smaller
than that currently adopted the origin of this source can be interpreted within
DFP-scenario in terms of the subcritical initial field almost without any additional
assumptions.

 \begin{table}
  \caption{Parameters of long-period pulsars inferred from DFP-scenario}
   \label{tab-2}
    \begin{tabular}{lcc}
    \hline
Name   &  $<\dmf>/<\dmf_{\rm x}>$   & $B_0/B_{\rm cr}$  \\
     \hline
J170006-4157   &  1 & 0.08  \\
0352+309/X~Per &  1 & 0.3  \\
J1037.5-5647   &  1 & 0.3  \\
J2239.3+6116   &  1 & 0.1 \\
  \noalign{\smallskip}
  \hline
   \noalign{\smallskip}
J0103.6-7201   &  0.2 -- 1  & 1 \\
J0146.9+6121/V831\,Cas & 0.6 & 1  \\
  \noalign{\smallskip}
  \hline
   \noalign{\smallskip}
2S\,0114+650$^*$   & 0.03 & 1  \\
    \hline
   \end{tabular}
\newline
$^*$~See, however, discussion in the text.
  \end{table}

   \section{Discussion}\label{discussion}

The main result of this paper is that the problem posed by the existence of relatively
bright persistent long-period X-ray pulsars can be solved by an incorporation of the
subsonic propeller state into the evolutionary tracks of neutron stars in massive close
binaries computed by \citet{Urpin-etal-1998}. This solution implies that the neutron
stars in these systems undergo spherical accretion and their initial magnetic field is
subcritical. The spin periods of the neutron stars in this case are close to the
equilibrium period expressed by Eq.~(\ref{peqsph}) provided the relative velocity
between the star and surrounding gas is $V_{\rm rel} \sim 400-800\,{\rm km\,s^{-1}}$.
The only exception is the neutron star in 2S\,0114+650 whose equilibrium period for a
reasonable estimate of $V_{\rm rel}$ is by an order of magnitude smaller than $10^4$\,s.
It should be, however, noted that the association of the $10^4$\,s pulsations with the
spin period of the neutron star in this system remains so far controversial.

The evolutionary scenario presented seriously weakens the hypothesis that the neutron
stars in the long period pulsars are magnetar descendants. In particular, it is shown
that the interpretation of the long-period pulsar 2S\,0114+650 within this hypothesis is
not possible unless an assumption about a significantly (by a factor of 250) weaker wind
in a previous epoch is invoked. Furthermore, this hypothesis implies that magnetars form
during the first supernova explosion in a massive binary system, which is not consistent
with the current views on the process of magnetar formation. At the same time the above
arguments are not compelling to exclude the magnetar hypothesis completely and
therefore, a brief discussion about possible appearances of magnetars in close binaries
appears to be rather reasonable.

The main evolutionary stages of a magnetar with an initial magnetic field of $B_0 \sim 2
\times 10^{15}$\,G ($\mu_0 = 10^{33}~\mu_{33}\,{\rm G\,cm^3}$) in a binary system are as
follows \citep[for a discussion see, e.g.,][and references therein]{Ikhsanov-2001b}. The
{\it ejector} stage, which lasts
  \be\label{taua}
\tau_{\rm a} \sim 10^3\ \mu_{33}^{-1}\ I_{45}\ \dmf_{15}^{-1/2}\ V_8^{-1/2}\ {\rm yr},
 \ee
and ends as the spin period of the neutron star is
  \be\label{pmd}
P_{\rm s} = P_{\rm md} \sim 53\ \mu_{33}^{1/2}\ \dmf_{15}^{-1/4}\ V_8^{-1/4}\ {\rm s},
 \ee
where $V_8=V_{\rm rel}/10^8\,{\rm cm\,s^{-1}}$.

The parameters of the {\it propeller} stage depend on the geometry of the accretion flow
beyond the neutron star's magnetospheric boundary. If the magnetospheric radius of the
star is smaller than the so called circularization radius, which is defined as
 \be\label{rcirc}
r_{\rm circ} = \frac{\dot{J}^{2}}{\dmf^{2} GM},
 \ee
where
   \be\label{aar}
\dot{J}=\xi \dot{J}_{0} = \frac{1}{4} \xi \dmf \Omega_{\rm orb} r_{\rm eff}^{2}
   \ee
is the rate of accretion of angular momentum, the formation of a disc would be expected.
Combining Eqs.~(\ref{rcirc}) and (\ref{aar}) and solving the result for $V_{\rm rel}$
one finds the condition for a disc formation as
 \be\label{vrel}
V_{\rm rel} \la V_{\rm cr} \simeq 75\ \xi_{0.2}^{1/4}\ \mu_{32}^{-1/14}\ m^{11/28}\
P_{250}^{-1/4}\ \dmf_{15}^{1/28}\ {\rm km\,s^{-1}}.
 \ee
It is taken into account here that the magnetic field of a magnetar on a time scale of
$\sim 10^3$\,yr decreases by almost an order of magnitude \citep[see,
e.g.,][]{Colpi-etal-2000}. If this condition is satisfied the spin period of the neutron
star during the propeller stage decreases to a value of
  \be\label{pcdd}
P_{\rm s} \sim P_{\rm cd}^{\rm (d)} \simeq 400\ \kappa_{0.5}^{3/2}\ \mu_{32}^{6/7}\
m^{-5/7}\ \dmf_{15}^{-3/7}\ {\rm s},
   \ee
on a time scale of
  \be
\tau_{\rm c}^{\rm (d)} \simeq 2 \times 10^4\ \kappa_{0.5}\ \mu_{32}^{-3/7}\
\dmf_{15}^{-11/14}\ m^{-8/7}\ \left(\frac{V_{\rm cr}}{75\,{\rm
km\,s^{-1}}}\right)^{1/2}\ {\rm yr},
 \ee
and the magnetar switches its state to an accretor. Here $\kappa_{0.5}=\kappa/0.5$.

If the plasma flow beyond the magnetospheric boundary has a spherical geometry the
duration of the supersonic propeller state would be
  \be\label{tauc}
\tau_{\rm c}^{\rm (sp)} \simeq 6 \times 10^4\ I_{45}\ \mu_{32}^{-1}\ \dmf_{15}^{-1/2}\
V_8^{-3/2}\ {\rm yr}.
 \ee
The spin period of the star during this time increases to a value of
  \be\label{pcdsp}
P_{\rm s} \la P_{\rm cd}^{\rm (sp)} \simeq 10^3\ \mu_{33}^{6/7}\ m^{-5/7}\
\dmf_{15}^{-3/7}\ {\rm s}.
   \ee
As the star switches to the subsonic propeller state its spin period increases on a time
scale of
 \be
\tau_{\rm d} \sim 200\ \mu_{32}^{-8/7}\ I_{45}\ m^{2/7}\ \dmf_{15}^{-3/7}\ {\rm yr},
 \ee
to a value of
  \be\label{pbr}
P_{\rm br} \simeq 1.1 \times 10^4\ \mu_{32}^{6/7}\ \dmf_{15}^{-5/7}\ m^{-4/21}\ {\rm s}.
 \ee

The above estimates suggest that the spin period of magnetars accreting material from a
disc unlikely exceeds $500 L_{35}^{-3/7}$\,s, where $L_{35}$ is the X-ray luminosity of
the source expressed in units of $10^{35}\,{\rm erg\,s^{-1}}$. If the time scale of the
magnetic field decay of magnetars exceeds $3 \times 10^4$\,yr a probability of
observational identification of these stars in the magnetar stage is not zero. On the
other hand, it appears to be rather small as the magnetic field of an accreting star
decays more rapidly than that of an isolated one \citep[for a discussion, see,
e.g.,][and references therein]{Urpin-etal-1998}. The spin period of magnetars in the
accretor state evolves to the equilibrium period expressed by Eq.~(\ref{peqd}) and the
time scale of its evolution is determined by the decay rate of the stellar magnetic
field.

The magnetars undergoing spherical accretion can be spun-down to a period of $\sim 2
\times 10^4 L_{35}^{-5/7}$\,s. However, the spin-down time scale in this case proves to
be comparable with the upper limits to the characteristic time of the magnetic field
decay. Therefore, a possibility to observe an accretion-powered magnetar undergoing
spherical accretion is almost negligible.

Finally, the appearance of close binaries containing magnetar descendants of an age
$\sim t_{\rm ms}$ would be similar to that of close binaries in which the initial field
of the neutron star was under the critical value. In contrast, the evolutionary tracks
and appearance of neutron stars undergoing disc and spherical accretion differ
significantly.

\section*{Acknowledgements}

I would like to thank Malvin Ruderman and James E. Pringle for useful discussion and an
anonymous referee for very stimulative comments. I acknowledge the support of the
European Commission under the Marie Curie Incoming Fellowship Program. The work was
partly supported by Russian Foundation of Basic Research.


\begin{thebibliography}{}
\bibitem[Aab \& Bychkova(1984)]{Aab-Bychkova-1984}
 Aab, O.E., Bychkova, L.V. 1984, Sov. Ast. Letters, 9, 313
\bibitem[Anzer et~al(1987)]{Anzer-etal-1987}
 Anzer, U., B\"orner, G., Monaghan, J.J. 1987, A\&A 176, 235
\bibitem[Arons \& Lea(1976)]{Arons-Lea-1976}
 Arons, J., Lea, S.M. 1976, ApJ, 207, 914
\bibitem[Bernacca \& Bianchi(1981)]{Bernacca-Bianchi-1981}
 Bernacca, P.L., Bianchi, L. 1981, A\&A, 94,345
\bibitem[Bhattacharya \& van den Heuvel(1991)]{Bhattacharya-van-den-Heuvel-1991}
 Bhattacharya, D., van den Heuvel, E.P.J. 1991, Phys. Rep., 203, 1
\bibitem[Coburn et~al.(2001)]{Coburn-etal-2001}
 Coburn, W., Heindl, W.A., Gruber, D.E., et~al. 2001, ApJ, 552, 738
\bibitem[Colpi et~al.(2000)]{Colpi-etal-2000}
 Colpi, M., Geppert, U., Page, D. 2000, ApJ, 529, L29
%\bibitem[Corbet et~al.(1999)]{Corbet-etal-1999}
% Corbet, R.H.D., Finley, J.P., Peele, A.G. 1999, ApJ, 511, 876
\bibitem[Crowther, Lennon \& Walborn(2006)]{Crowther-Lennon-Walborn-2006}
 Crowther, P.A., Lennon,D.J., Walborn, N.R. 2006, A\&A, 446, 279
\bibitem[Davidson \& Ostriker(1973)]{Davidson-Ostriker-1973}
 Davidson, K., Ostriker, J.P. 1973, ApJ, 179,585
\bibitem[Davies et~al.(1979)]{Davies-etal-1979}
 Davies, R.E., Fabian, A.C., Pringle, J.E. 1979, MNRAS, 186, 779
\bibitem[Davies \& Pringle(1981)]{Davies-Pringle-1981}
 Davies, R.E., Pringle, J.E. 1981, MNRAS, 196, 209
\bibitem[Delgado-Marti et~al.(2001)]{Delgado-Marti-etal-2001}
 Delgado-Marti,H., Levine, A.M., Pfahl, E., Rappaport, A. 2001, ApJ, 546, 455
\bibitem[Elsner \& Lamb(1976)]{Elsner-Lamb-1976}
 Elsner, R.F., Lamb, F.K. 1976, Nature, 262, 356
\bibitem[Elsner \& Lamb(1984)]{Elsner-Lamb-1984}
 Elsner, R.F., Lamb, F.K. 1984, ApJ, 278, 326
\bibitem[Fullerton et~al.(2006)]{Fullerton-Massa-Prinja-2006}
 Fullerton, A.W., Massa, D.L., Prinja, R.K. 2006, ApJ, 637, 1025
\bibitem[Ghosh \& Lamb(1978)]{Ghosh-Lamb-1978}
 Ghosh, P., Lamb, F.K. 1978, ApJ, 223, L83
\bibitem[Gosling et~al.(1991)]{Gosling-etal-1991}
 Gosling, J.T., Thomsen, M.F., Bame, S.J., et~al. 1991, J. Geophys. Res. 96, 14097
\bibitem[Haberl(1994)]{Haberl-1994}
 Haberl, F. 1994, A\&A, 283, 175
\bibitem[Haberl et~al.(1998)]{Haberl-etal-1998}
 Haberl, F., Angelini, L., Motch, C., White, N.E. 1998, A\&A, 330, 189
\bibitem[Haberl \& Pietsch(2005)]{Haberl-Pitsch-2005}
 Haberl, F., Pitsch, W. 2005, A\&A, 438, 211
\bibitem[Haberl \& Sasaki(2000)]{Haberl-Sasaki-2000}
 Haberl, F., Sasaki, M. 2000, A\&A, 359, 753
\bibitem[Heger et~al.(2003)]{Heger-etal-2003}
 Heger, A., Fryer, C.L., Woosley, S.E., Langer, N., Hartmann, D.H. 2003, ApJ, 591, 288
\bibitem[Hall et~al.(2000)]{Hall-etal-2000}
 Hall, T.A., Finley, J.P., Corbet, R.H.D., Thomas,R.C. 2000, ApJ, 536, 450
\bibitem[Iben, Tutukov \& Yungelson(1995)]{Iben-Tutukov-Yungelson-1995}
 Iben, I. Jr., Tutukov, A.V., Yungelson, L.R. 1995, ApJS, 100, 217
\bibitem[Ikhsanov(2001a)]{Ikhsanov-2001a}
 Ikhsanov, N. 2001a, A\&A, 368, L5
\bibitem[Ikhsanov(2001b)]{Ikhsanov-2001b}
 Ikhsanov, N.R. 2001b, A\&A, 375, 944
\bibitem[Ikhsanov(2002)]{Ikhsanov-2002}
 Ikhsanov, N. 2002, A\&A, 381, L61
\bibitem[Ikhsanov \& Pustilnik(1996)]{Ikhsanov-Pustilnik-1996}
 Ikhsanov, N., Pustilnik, L.A.  1996, A\&A, 312, 338
\bibitem[in 't Zand et~al.(2000)]{in-'t-Zand-etal-2000}
 In 't Zand, J.J.M., Halpern, J., Eracleous, M., McCollough, M., Augusteijn, T.,
   Remillard, R.A., Heise, J. 2000, A\&A, 361, 85
\bibitem[in 't Zand et~al.(2001)]{in-'t-Zand-etal-2001}
 In 't Zand, J.J.M., Swank, J., Corbet, R.H.D., Markwardt, C.B. 2001, A\&A, 380, L26
\bibitem[Illarionov \& Sunyaev(1975)]{Illarionov-Sunyaev-1975}
 Illarionov, A.F., Sunyaev, R.A. 1975 A\&A, 39, 185
\bibitem[Koenigsberger et~al.(1983)]{Koenigsberger-etal-1983}
 Koenigsberger, G., Swank, J.H., Szymkowiak, A.E., White, N.E. 1983, ApJ, 268, 782
\bibitem[Koenigsberger et~al.(2006)]{Koenigsberger-etal-2006}
 Koenigsberger, G., Georgiev, L., Moreno, E., Richer, M.G., Toledano, O., Canalizo, G., Arrieta, A.
 2006, astro-ph/0608226~v1
\bibitem[Lamb et~al.(1977)]{Lamb-etal-1977}
 Lamb, F.K., Fabian, A.C., Pringle, J.E., \& Lamb, D.Q. 1977, ApJ, 217, 197
\bibitem[Li \& van den Heuvel(1999)]{Li-van-den-Heuvel-1999}
 Li, X.-D., van den Heuvel, E.P.J. 1999, ApJ, 513, L45
\bibitem[Lipunov(1992)]{Lipunov-1992}
 Lipunov, V.M. 1992, Astrophysics of neutron stars,
 Springer-Verlag, Heidelberg
% \bibitem[Liu et~al.(2000)]{Liu-etal-2000}
%  Liu, Q.Z., van Paradijs, P., van den Heuvel, E.P.J. 2000, A\&AS, 147, 25
% \bibitem[Liu et~al.(2005)]{Liu-etal-2005}
%  Liu, Q.Z., van Paradijs, P., van den Heuvel, E.P.J. 2005, A\&A, 442, 1135
\bibitem[Lynden-Bell \& Pringle(1974)]{Lynden-Bell-Pringle-1974}
 Lynden-Bell, D., Pringle, J.E. 1974, MNRAS, 168, 603
%\bibitem[Matsuda et~al.(1991)]{Matsuda-etal-1991}
% Matsuda, T., Sekino, N., Sawada, K., Shima, E., Livio, M., Anzer, U., B\"orner, G.
% 1991, A\&A, 248, 301
\bibitem[Marlborough(1982)]{Marlborough-1982}
 Marlborough, J.M. 1982, in Be stars, eds. M.\,Jaschek, H.G.\,Groth, Reidel, p.\,361
\bibitem[Meynet et~al.(1994)]{Meynet-etal-1994}
 Meynet, G., Maeder, A., Schaller, G., Schearer, D., Charbonel, C. 1994, A\&AS, 103, 97
\bibitem[Motch et~al.(1997)]{Motch-etal-1997}
 Motch, C., Haberl, F., Dennerl, K., Pakull, M., Janot-Pacheco, E. 1997, A\&A, 323, 853
% \bibitem[Okazaki \& Negueruela(2001)]{Okazaki-Negueruela-2001}
%  Okazaki, A.T., Negueruela, I. 2001, A\&A, 377, 161
% \bibitem[Okazaki et~al.(2002)]{Okazaki-etal-2002}
%  Okazaki, A.T., Bate, M.R., Ogilvie, G.I., Pringle, J.E. 2002, MNRAS, 337, 967
\bibitem[Petrovic et~al.(2005)]{Petrovic-etal-2005}
 Petrovic, J., Langer, N., Yoon, S.-C., Heger, A. 2005, A\&A, 435, 247
\bibitem[Popov \& Prokhorov(2006)]{Popov-Prokhorov-2006}
 Popov, S.B., Prokhorov, M.E. 2006, MNRAS, 367, 732
\bibitem[Price \& Rosswog(2006)]{Price-Rosswog-2006}
 Price, D.J., Rosswog, S. 2006, Science, 312, 719
\bibitem[Pringle \& Rees(1972)]{Pringle-Rees-1972}
 Pringle, J.E., Rees, M.J. 1972, A\&A, 21, 1
% \bibitem[Ragizova \& Popov(2005)]{Raguzova-Popov-2005}
%  Raguzova, N.V., Popov, S.B. 2005, A\&AT, 24, 151
\bibitem[Reig et~al.(1996)]{Reig-etal-1996}
 Reig, P., Chakrabarty, D., Coe, M.J., et~al. 1996, A\&A, 311, 879
\bibitem[Reig \& Roche(1999)]{Reig-Roche-1999}
 Reig, P., Roche, P. 1999, MNRAS, 306, 95 \& 100
% \bibitem[Reig et~al.(2004)]{Reig-etal-2004}
% Reig, P., Negueruela, I., Fabregat, J., Chato, R., Blay, P., Mavromatakis, F.
%  2004, A\&A, 421, 673
\bibitem[Ruffert(1999)]{Ruffert-1999}
 Ruffert, M. 1999, A\&A, 346, 861
% \bibitem[Shapiro \& Lightman(1976)]{Shapiro-Lightman-1976}
%  Shapiro, S.L., Lightman, A.P. 1976, ApJ, 204, 555
\bibitem[Snow(1981)]{Snow-1981}
 Snow, T.P. 1981, ApJ, 251, 139
\bibitem[Stella, White \& Rosner(1986)]{Stella-White-Rosner-1986}
 Stella, L., White, N.E., Rosner, R. 1986, ApJ, 308, 669
% \bibitem[Sunyaev \& Shakura(1977)]{Sunyaev-Shakura-1977}
%  Sunyaev, R.A., Shakura, N.M. 1977, SvAL, 3, 138
\bibitem[Taam \& Fryxell(1988)]{Taam-Fryxell-1988}
 Taam R.E., Fryxell B.A., 1988, ApJ 327, L73
\bibitem[Thompson \& Duncan(1993)]{Thompson-Duncan-1993}
 Thompson, C., Duncan, R.C. 1993, ApJ, 408,194
\bibitem[Thompson \& Murray(2001)]{Thompson-Murray-2001}
 Thompson, C., Murray, N. 2001, ApJ, 560, 339
\bibitem[Torii et~al.(1999)]{Torii-etal-1999}
 Torii, K., Sugizaki, M., Kohmura, T., Endo, T., Nagase, F. 1999, ApJ, 523, L65
\bibitem[Urpin et~al.(1998)]{Urpin-etal-1998}
 Urpin, V, Konenkov, D., Geppert, U. 1998, MNRAS, 299, 73
% \bibitem[van den Heuvel(1977)]{van-den-Heuvel-1977}
%  van den Heuvel, E.P.J. 1977, Ann. N.Y. Acad. Sci., 302, 15
% \bibitem[van den Heuvel(1981)]{van den Heuvel-1981}
%  van den Heuvel, E.P.J. 1981, SSRv, 30, 623
\bibitem[Vink et~al.(2000)]{Vink-de-Koter-Lamers-2000}
 Vink, J.S., de Koter, A., Lamers, H.J.G.L.M. 2000, A\&A, 362, 295
\bibitem[Wang(1981)]{Wang-1981}
 Wang, Y.-M. 1981, A\&A, 102, 36
\bibitem[Wheeler et~al.(2000)]{Wheeler-etal-2000}
 Wheeler, J.C., Yi, I., Hoflich, P., Wang, L. 2000, ApJ, 537, 810
% \bibitem[Woods \& Thompson(2004)]{Woods-Thompson-2004}
%  Woods, P.M., Thompson, C. 2004, in ``Compact Stellar X-ray Sources'',
% Lewin, H.G., van der Klis, M., eds., preprint (astro-ph/0406133)
\bibitem[Yamauchi et~al.(1990)]{Yamauchi-etal-1990}
 Yamauchi, S., Asaoka, I., Kawada, M., Koyama, K., Tawara, Y. 1990, PASJ, 42, L53
\bibitem[Ziolkowski(2002)]{Ziolkowski-2002}
 Ziolkowski, J. 2002, Mem. Soc. Astron. Ital., 73, 1038
\end{thebibliography}
\end{document}